\documentclass{article}
\usepackage{amssymb}

\begin{document}

\title{The Nature of Dark Matter}
\author{A. A. Kirillov \\
\emph{Institute for Applied Mathematics and Cybernetics} \\
\emph{10 Uljanova Str., Nizhny Novgorod, 603005, Russia}}
\date{}
\maketitle

\begin{abstract}
The observed strong dark-to-luminous matter coupling \cite{S04,Core,CE}
suggests the existence of a some functional relation between visible and DM
sources which leads to biased Einstein equations. We show that such a bias
appears in the case when the topological structure of the actual Universe at
very large distances does not match properly that of the Friedman space. We
introduce a bias operator $\rho _{DM}$ $=\widehat{B}\rho _{vis}$ and show
that the simple bias function $b=1/(4\pi r_{0}r^{2})\theta \left( r-r_{\max
}\right) $ (the kernel of $\widehat{B}$) allows to account for all the
variety of observed DM halos in astrophysical systems. In galaxies such a
bias forms the cored DM distribution with the radius $R_{C}\sim R_{opt}$
(which explains the recently observed strong correlation between $R_{C}$ and 
$R_{opt}$ \cite{S04}), while for a point source it produces the logarithmic
correction to the Newton's potential (which explains the observed flat
rotation curves in spirals). Finally, we show that in the theory suggested
the galaxy formation process leads to a specific variation with time of all
interaction constants and, in particular, of the fine structure constant.
\end{abstract}

\newpage

\section{Introduction}

The existence of Dark Matter (DM) has been long known \cite{Zw}. It
represents the most mysterious phenomenon of our Universe which still did
not find a satisfactory explanation in modern physics. While more than $90\%$
of matter of the Universe has a non-baryonic dark form, lab experiments show
no evidence for the existence of such matter. The success of (Lambda) Cold
Dark Matter (CDM) models in reproducing the large-scale structure is
accompanied with a failure in describing the Universe on smaller scales.
Indeed, it is now well established that in galaxies the Dark Matter (DM)
density shows an inner core, i.e. a central constant density region (e.g.,
see Refs. \cite{Core} for spirals and Refs. \cite{CE} for ellipticals and
references therein). Such a feature is in clear conflict with $\Lambda -$CDM
models which predict the presence of cusps ($\rho _{DM}\sim 1/r$) in the
inner regions of galaxies \cite{NFW} (see however a more positive view in
Ref. \cite{Pr}). The situation is somewhat better for the Milgrom's
algorithm \cite{mil} MOND (Modified Newtonian Dynamics). However, the
existence of a very strong correlation between the core radius size $R_{C}$
and the stellar exponential scale length $R_{D}$ (or the optical radius $%
R_{opt}$\footnote{$R_{opt}$ is the radius encompassing 83\% of the total
luminosity of the galaxy. In the case of a (stellar) exponential thin disk $%
R_{opt}$ is 3.2 times the disk scale length $R_{D}$.}), $R_{C}\simeq 13\ (%
\frac{R_{D}}{5\ \mathrm{kpc}})^{1.05}\ \mathrm{kpc}$, e.g., see Ref. \cite%
{S04}, rules out MOND as well. Indeed, according to Milgrom's algorithm the
low acceleration regime triggers off at $R_{MOND}$, when the gravitation
acceleration $g=GM_{gal}/r^{2}$ drops below a fundamental acceleration $%
a_{0}\sim 2\times 10^{-8}cm/\,s^{2}$ (i.e., $R_{MOND}^{2}\sim GM_{gal}/a_{0}$%
), and in general the two parameters $R_{D}$ and $R_{MOND}$ are independent.
By other words there should exist galaxies in which either $R_{D}\ll $ $%
R_{MOND}$, or $R_{D}\gg $ $R_{MOND}$. And indeed an example of such a galaxy
has been recently presented in Ref. \cite{05}.

Thus we see that the modern theory of structure formation faces a rather
difficult situation. Main alternatives to CDM, worm DM and self-interacting
DM, seem to be ruled out by data on large scales (e.g., see Ref. \cite{Pr}
and references therein), while the distribution of DM in galaxies rules out
CDM \cite{S04,Core} and MOND \cite{05}.

The correlation between the core size $R_{C}$ and the optical size $R_{opt}$
in galaxies of different morphological type \cite{S04} points clearly out to
the presence of a very strong coupling between DM halos and baryons which
surely requires some new physics. We recall that such a strong
dark-to-luminous matter coupling (the so-called bias) is actually observed
on all scales (e.g., Ref. \cite{Pr,book}). In general this means the
existence of a functional dependence or the so-called bias relation $T_{\mu
\nu }^{DM}=F_{\mu \nu }(T_{\alpha \beta })$ between DM $T_{\mu \nu }^{DM}$
and the visible matter $T_{\mu \nu }$ sources. In the linear case the bias
can be expressed by 
\begin{equation}
T_{\mu \nu }^{DM}=\widehat{B}T_{\mu \nu }=\int_{x^{\prime }<x}B_{\mu \nu
}^{~\ \ \ \ \alpha \beta }\left( x,x^{\prime }\right) T_{\alpha \beta
}\left( x^{\prime }\right) d\Omega ^{\prime },  \label{br}
\end{equation}%
where to save the causality the integration should be taken over the
past-light-cone of the point $x$. In CDM models the bias relation appears as
a result of the nonlinear dynamics during the structure formation and
carries a nonlinear character, while on very large scales, where
inhomogeneities are still in the linear regime, such a bias should be viewed
as the result of a generation process of primordial perturbations or merely
as a result of the specific choice of initial conditions. In the present
paper we consider the simplest case, i.e., the isotropic and homogeneous
Universe with visible matter in the form of dust. Then the bias operator can
be expressed via a single function $B_{\mu \nu }^{\alpha \beta }\left(
x,x^{\prime }\right) =\left( \delta _{\mu }^{\alpha }\delta _{\nu }^{\beta
}+\delta _{\nu }^{\alpha }\delta _{\mu }^{\beta }\right) b\left(
t,x-x^{\prime }\right) $. Moreover, in such a case the bias function $%
b\left( t,x-x^{\prime }\right) $ can be fixed from observational data, e.g.,
for Fourier transforms the bias relation (\ref{br}) gives 
\begin{equation}
T_{\mu \nu }^{DM}\left( t,k\right) =B\left( t,k\right) \ T_{\mu \nu
}^{vis}\left( t,k\right)
\end{equation}%
which allows to find empirically the bias operator $\widehat{B}_{emp}$. And
it is quite obvious that the empirical bias operator $\widehat{B}_{emp}$ (in
virtue merely of its definition) will perfectly describe DM effects at very
large scales (i.e. in the region of linear perturbations). Any actual
specific source of DM (to fit observations) should reproduce properties of
the bias operator $\widehat{B}_{emp}$ in details.

The bias relation allows to re-write the Einstein equations in the
equivalent form%
\begin{equation}
R_{\mu \nu }-\frac{1}{2}g_{\mu \nu }R=8\pi G\left( T_{\mu \nu }+F_{\mu \nu
}(T_{\alpha \beta })\right) .  \label{Ee}
\end{equation}%
Now we can forget about the origin of the bias and study straightforwardly
equations in the form (\ref{Ee}). The advantage is that equations (\ref{Ee})
do not imply the existence of any actual DM source. Therefore, with the same
success we can interpret (\ref{Ee}) as a specific modification of gravity.
Most of modifications suggested (e.g., see Refs \cite{mil,B,log}) can be
reformulated in the form (\ref{Ee}). In particular, for a point mass at rest
equations (\ref{Ee}) lead to a Modified Newton's law 
\begin{equation}
\phi =-\frac{GM_{0}}{r}\left( 1+f\left( t,r\right) \right) ,  \label{mnl}
\end{equation}%
where in general the correction $f\left( t,r\right) $ depends also on the
position of the point source in space. We also note that such a modification
can be equally interpreted as a specific "renormalization" of the
gravitational constant $G\rightarrow G\left( 1+\widehat{B}\right) $ (e.g.,
see Refs. \cite{Ob,KT02}).

In the present paper we discuss the bias relation which appears in the case
when the topological structure of the physical space (i.e., of the Universe)
does not match properly that of the Friedman space. It was demonstrated
recently (e.g., see Refs. \cite{KT02,K03}) that in this case the standard
Newton's law violates (there exist a range of scales $r_{0}<r<r_{\max }$ in
which the gravitational potential has the logarithmic behavior, i.e., $%
f\left( t,r\right) =r/r_{0}\ln r$). We show that the simple bias predicted
in Refs. \cite{KT02,K03} $b=1/4\pi r_{0}\left\vert r-r^{\prime }\right\vert
^{2}\theta \left( r-r_{\max }\right) $ gives a rather good qualitative
agreement with the observed picture of the Universe at smaller scales. In
particular, such a bias allows to relate together a number of observational
facts. Namely, the asymptotically flat rotation curves of spiral galaxies 
\cite{PSS} (which indicate that starting from some length scale $r_{0}$ the
gravity force behaves as $1/r$), the cored distribution of DM density in
galaxies \cite{Core,CE}, the observed very strong correlation between $R_{C}$
and $R_{D}$ \cite{S04}, and the fractal behavior in the distribution of
galaxies (which has the dimension $D\simeq 2$ and is observed at least up to 
$200Mpc$ \cite{fr}). In the view of the modification of the Newton's law (%
\ref{mnl}) the last fact indicates that the maximal scale $r_{\max }$ after
which the standard gravity law restores (e.g., it becomes $F\sim 1/r^{2}$
again) should be $r_{\max }>200Mpc$ \cite{k02}).

All these facts are well established and are beyond doubts. There were some
debates in the literature about the fractal distribution of galaxies \cite%
{fr2}. However, the test for the fractality is rather simple, e.g., if we
consider any galaxy, surround it with a sphere of a radius $R$, and count
for the number of galaxies $N\left( R\right) $ within the radius $R$, we
find the law $N\left( R\right) \sim R^{D}$. And the value $D$ is, in turn,
not sensible to small perturbations of the galaxy distribution which may
appear due to uncertainties in distances\footnote{%
The misunderstanding may appear if one performs an averaging over the
central position of the sphere in space. In this case one gets nothing but
the trivial result $D\approx 3$.}. Moreover, the large-scale structure,
e.g., the existence of huge ($\sim 100-200Mpc$) voids with no galaxies
inside and thin filled with galaxies walls ($\sim 1-5Mpc$), is quite
consistent with $D\simeq 2$. Thus, it is safe to accept the fractal picture,
at least up to $200$ $Mpc$.

\section{Origin of the bias}

In the present section we show that a non-trivial topological structure of
the physical space can quite naturally give rise to the origin of the bias 
\cite{KT02,K03}. Indeed, in considering astrophysical systems we use an
extrapolation of spatial relationships which are well-tested on considerably
smaller scales. Therefore, if the topological structure of the actual
Universe at very large distances does not match properly that of the
Friedman space (the open, flat, or closed model) we naturally should observe
some discrepancy. To describe such a discrepancy we first consider an
example from solid state physics.

Consider a medium of a low density at very small temperatures. From the
thermodynamics we know that most of systems at a sufficiently small
temperature acquire a crystal structure. However, in actual systems such a
crystal has never an ideal character but includes different distortions.
Moreover, when a system has a rather low density and the rate of freezing is
rapid enough, such a system will include considerable voids and the spatial
distribution of particles in the system acquires, in turn, quite irregular
character. Elementary excitations (or quasiparticles, e.g., electrons of the
conductivity, phonons, etc.) in the given system do exist only within the
crystal and from their point of view the physical space (the crystal)
possesses a rather non-trivial topological structure. From the mathematical
standpoint the non-trivial topological structure can be accounted for as
follows.

Consider a volume $V$ in $R^{3}$ occupied with a system and let $H$ be the
Hilbert space for a free particle (the space of functions on $V$). Let $%
\left\{ g_{k}\left( x\right) \right\} $ ( $x\in V$) be an arbitrary basis in 
$H$. Physically, the basis represents a set of eigenvectors for a complete
set of observables. E.g., for a scalar (without the spin) particle we can
use the coordinate representation (i.e., $g_{k}\left( x\right) =\delta
\left( x_{k}-x\right) $ is the set of eigenvectors for the position operator 
$\widehat{X}g_{k}=x_{k}g_{k}$, $x_{k}\in V$) or the momentum representation (%
$g_{k}\left( x\right) =(V)^{-1/2}\exp \left( ikx\right) $, so that $\widehat{%
P}g_{k}=kg_{k}$). The basis is supposed to be normalized $%
(g_{k},g_{p})=\delta _{kp}$ and complete $\sum g_{k}^{\ast }\left( x\right)
g_{k}\left( x^{\prime }\right) =\delta \left( x-x^{\prime }\right) $, where $%
x$, $x^{\prime }\in V$. The fact that our system has an irregular
distribution in $V$ (i.e., $V$ includes also voids) means that some states
in $H$ cannot be physically realized for particles of the system (at least
for small temperatures when the structure of the crystal does not change).
Thus, we have to restrict the space of states $H$ to the space of physically
admissible states $H_{phys}=\widehat{K}H$, where $\widehat{K}=$ $(\widehat{K}%
)^{2}$ is a projection operator. In the basis of eigenvectors the projection
operator $\widehat{K}$ takes the diagonal form $(f_{i},\widehat{K}%
f_{k})=K_{ik}=N_{k}\delta _{ik}$ with eigenvalues $N_{k}=0,1$. Thus, an
arbitrary (but physically realizable) state of a particle is biased and can
be presented as $\psi _{phys}=\widehat{K}^{1/2}\psi =\sum \sqrt{N_{k}}%
a_{k}f_{k}\left( x\right) $. Thus we see that topological structure of the
system is described by the bias (projection) operator $\widehat{K}$. In
particular, all physical observables acquire the structure $\widehat{O}%
_{phys}=\widehat{K}^{1/2}~\widehat{O}~\widehat{K}^{1/2}$, while the physical
space $V_{phys}$ of the system represents the space of eigenvalues $x_{k}\in
V_{phys}$ of the biased position operator of a particle $\widehat{X}_{phys}=%
\widehat{K}^{1/2}~\widehat{X}~\widehat{K}^{1/2}$.

In the example described the bias operator is diagonal in the coordinate
representation (i.e., $N_{k}=0$, when $x_{k}$ belongs to voids and $N_{k}=1$
as $x_{k}$ belongs to the crystal). However, we can also consider a more
general case when $\widehat{K}$ and $\widehat{X}$ do not have common
eigenvectors (i.e., $[\widehat{K}$, $\widehat{X}]\neq 0$). In the last case
the spatial structure of the crystal remains unspecified. This means that in
such a system the position operator cannot be a good observable (at least
while the topological structure of the system conserves, i.e., $K_{ik}=const$%
, which is always fulfilled at sufficiently small temperatures). We also
note that from the point of view of the mathematical coordinate space (i.e., 
$R^{3}$) the space $H_{phys}$ is not complete, i.e., $\sum N_{k}f_{k}^{\ast
}\left( x\right) f_{k}\left( x^{\prime }\right) $ $=$ $K(x,x^{\prime })$ $=$ 
$\widehat{K}^{1/2}\delta \left( x-x^{\prime }\right) \widehat{K}^{1/2}~~\neq 
$ $\delta \left( x-x^{\prime }\right) $. Thus, we see that the function $%
K(x,x^{\prime })$ plays here the role of the delta function. And only in the
case when both $\widehat{K}$ and $\widehat{X}$ can be diagonalized
simultaneously the biased delta function $K(x,x^{\prime })$ reduces to the
ordinary delta function $K(x,x^{\prime })=\delta \left( x-x^{\prime }\right)
\theta (x,V_{phys})$, where $\theta (x,V_{phys})$ is a characteristic
function, i.e., $\theta (x,V_{phys})=0$ as $x\notin V_{phys}$ and $\theta
(x,V_{phys})=1$as $x\in V_{phys}$.

At very low temperatures the structure of the crystal conserves. This means
that the projection operator $\widehat{K}$ represents an integral of motion
(commutes with the Hamiltonian of the system). Therefore, we can state that
elementary excitations (quasi-particles) represent eigenvectors for the
projection operator i.e., the wave function of an excitation can be expanded
as $\psi _{phys}=\sum \sqrt{N_{k}}a_{k}f_{k}\left( x\right) $, while the
energy of the system can be represented as%
\begin{equation}
E=\sum N_{k}\varepsilon \left( k\right) a_{k}^{+}a_{k},
\end{equation}%
where $\varepsilon \left( k\right) $ is the energy of a quasi-particle.
Thus, we see that the non-trivial topological structure of the system
defines the measure (i.e., the density of degrees of freedom) which can be
accounted for by the formal substitution%
\begin{equation}
\sum_{k}\rightarrow \sum_{k}N_{k}  \label{sb}
\end{equation}%
(indeed, the algebra of physical observables modifies as $A=BC$ $\rightarrow 
$ $A_{phys}=B_{phys}C_{phys}=\widehat{K}^{1/2}~B~\widehat{K}C~\widehat{K}%
^{1/2}$ and $(B~\widehat{K}C)_{ij}=\sum_{k}N_{k}B_{ik}C_{kj}$). Any point
source for quasiparticles is always biased (as compared to the simple
topology case), i.e., acquires a specific distribution in $R^{3}$%
\begin{equation}
\delta \left( x-x^{\prime }\right) \rightarrow K(x,x^{\prime })=\widehat{K}%
^{1/2}\delta \left( x-x^{\prime }\right) \widehat{K}^{1/2}~  \label{ds}
\end{equation}%
which reflects the topological structure of the system (the discrepancy
between $V_{phys}$ and $V$). In particular, the actual physical volume
occupied by the crystal is given by $V_{phys}$ $=$ $\int_{V}K\left(
x,x^{\prime }\right) d^{3}xd^{3}x^{\prime }$ $\neq V$.

The above construction generalizes straightforwardly onto relativistic
particles. In a curved space the one-particle Hilbert space is not well
defined, for particles are actually not free. This means that in general
there is no such an observable as the position operator $\widehat{X}$ or the
momentum $\widehat{P}$ to classify quantum states. We recall the well-known
fact that even in the flat space the momentum of a particle can be
considered as a good operator, while the position operator is not. It can be
defined though by means of the Newton-Wigner construction \cite{NW}. Thus,
in this case the space of quantum states is constructed as follows.

Consider an arbitrary set of solution to the wave equation\footnote{%
If we require that the topological structure should be invariant under
conformal transformations, then we should set $m=0$ in (\ref{w}).} 
\begin{equation}
\left( \square +\frac{1}{6}R+m^{2}\right) f_{k}=0,  \label{w}
\end{equation}%
(where $\square f_{k}=\frac{1}{\sqrt{-g}}\partial _{\alpha }\left( \sqrt{-g}%
g^{\alpha \beta }\partial _{\beta }f_{k}\right) $) which obey the
normalization conditions 
\begin{equation}
\left( f_{k},f_{j}\right) =-\left( f_{k}^{\ast },f_{j}^{\ast }\right)
=\delta _{kj},\ \left( f_{k}^{\ast },f_{j}\right) =0\ ,  \label{ss}
\end{equation}%
and the scalar product is defined as (e.g., see Ref. \cite{br}) 
\begin{equation}
\left( f_{1},f_{2}\right) =i\int \left( f_{1}^{\ast }\left( x\right) \nabla
_{\mu }f_{2}\left( x\right) -f_{2}\left( x\right) \nabla _{\mu }f_{1}^{\ast
}\left( x\right) \right) \sqrt{-g}d\Sigma ^{\mu }.
\end{equation}%
Then the space of one-particle quantum states $H^{1}$ is defined as the
space of "positive frequency" solutions $\left\{ f_{k}\right\} $. And again
in simple cases a non-trivial structure of the physical space can be
accounted for by the fact that some of one-particle quantum states cannot be
physically realized, i.e., we should project the space of states $H^{1}$ to
the space of physically admissible states $H_{phys}^{1}=\widehat{K}H^{1}$.
In general the projection (bias) operator distinguishes a particular
(preferred) basis $\left\{ f_{k}\right\} $ in terms of which it can be
presented as\footnote{%
We note that the operator $K_{\phi }$ defined in (\ref{bb}) acts in the
space of fields $\phi \left( x\right) $. In the one-particle Hilbert space
it has the standard form $\widehat{K}=\sum N_{k}\left\vert
1_{k}\right\rangle \left\langle 1_{k}\right\vert $.} 
\begin{equation}
K_{\phi }\left( x,x^{\prime }\right) =\sum N_{k}\left( f_{k}\left( x\right)
f_{k}^{\ast }\left( x^{\prime }\right) -f_{k}^{\ast }\left( x\right)
f_{k}\left( x^{\prime }\right) \right) ,  \label{bb}
\end{equation}%
with eigenvalues $N_{k}=0,1$. Thus, physical fields can be defined as biased
fields%
\begin{equation}
\phi _{phys}=\widehat{K}_{\phi }^{1/2}\phi =\sum \sqrt{N_{k}}\left(
a_{k}f_{k}\left( x\right) +a_{k}^{+}f_{k}^{\ast }\left( x\right) \right)
\label{bmod}
\end{equation}%
and the nontrivial topological structure of space is reflected in the fact
that some modes never enter the expansion (\ref{bmod}) (i.e., for which $%
N_{k}=0$). And again any physical observable (i.e., every operator) can be
expressed as $\widehat{O}_{phys}=\widehat{K}^{1/2}~\widehat{O}~\widehat{K}%
^{1/2}$. E.g., in the case of a scalar field the mean value for the stress
energy tensor is biased as%
\begin{equation}
\left\langle n_{k}\left\vert T_{\alpha \beta }^{phys}\right\vert
n_{k}\right\rangle =\left\langle n_{k}\left\vert \widehat{K}^{1/2}T_{\alpha
\beta }\widehat{K}^{1/2}\right\vert n_{k}\right\rangle =\sum_{k}N_{k}\left(
1+2n_{k}\right) T_{\alpha \beta }\left[ f_{k}\left( x\right) ,f_{k}^{\ast
}\left( x\right) \right] ~,
\end{equation}%
where $T_{\alpha \beta }\left[ \phi ,\phi \right] $ is given by the bilinear
form 
\begin{equation}
T_{\alpha \beta }\left[ \phi ,\phi ^{\ast }\right] =\phi _{,\alpha }\phi
_{,\beta }^{\ast }-\frac{1}{2}g_{\alpha \beta }\left( g^{\mu \nu }\phi
_{,\mu }\phi _{,\nu }^{\ast }-m^{2}\phi \phi ^{\ast }\right)
\end{equation}%
and $\left\vert n_{k}\right\rangle =\prod \left( n_{k}!\right) ^{-1/2}\left(
a_{k}^{+}\right) ^{n_{k}}\left\vert 0\right\rangle $. The Green functions
for the physical scalar field (e.g., Feinman propagator $iG_{F}\left(
x,x^{\prime }\right) =\left\langle 0\left\vert T\phi _{phys}\left( x\right)
\phi _{phys}\left( x^{\prime }\right) \right\vert 0\right\rangle $) obey
formally the standard equation%
\begin{equation}
\left( \square +m^{2}\right) G_{F}\left( x,x^{\prime }\right) =\Delta \left(
x,x^{\prime }\right) \neq -\left( -g\right) ^{-1/2}\delta \left( x-x^{\prime
}\right) .  \label{M}
\end{equation}%
However the r.h.s. of this equation is not the delta function any more but
physical or biased delta function (\ref{ds}) (i.e., in terms of the
coordinate space it acquires an additional distribution in space $\Delta
\left( x,x^{\prime }\right) $ $=$ $-\widehat{K}_{\phi }^{1/2}$ $\left(
-g\left( x\right) \right) ^{-1/4}$ $\delta \left( x-x^{\prime }\right) $ $%
\left( -g\left( x^{\prime }\right) \right) ^{-1/4}$ $\widehat{K}_{\phi
}^{1/2}$). In this manner we see again that the role of the bias operator
(and that of the structure of the physical space) is the specification of
the density of degrees of freedom (\ref{sb}).

In conclusion of this section we point out to the two important facts. The
first is that the bias (\ref{bb}) includes a non-linear dependence on the
metric $g_{\alpha \beta }$ via the solution of Eq. (\ref{w}). And the second
is that the projection operator (bias) discussed above restricts strongly
the topological structure of the physical space. Indeed, by the construction
the projection $\widehat{K}=$ $(\widehat{K})^{2}$ means that the physical
space $V_{phys}$ represents a subspace in $R^{3}$ (i.e., $V_{phys}\subset
R^{3}$ or in cosmology it should represent a subspace of the Friedman
space). In the most general case however such an embedding may not exist. By
other words an arbitrary physical space (of an arbitrary topological
structure) cannot be projected to the Friedman space (or $R^{3}$) without
self-intersections (i.e., $\widehat{K}\neq $ $(\widehat{K})^{2}$) . This, in
turn, leads to a generalization of the bias operator (\ref{bb}) to the more
general case (e.g., see Refs. \cite{KT02,K03}) which naturally leads to the
generalized statistics of particles. From the formal standpoint such a
generalization is expressed by the fact that eigenvalues $N_{k}$ of the bias
operator $\widehat{K}$ can be arbitrary integer numbers $N_{k}=0,1,2,...$, ($%
N_{k}^{2}\neq N_{k}$ and $\widehat{K}\neq $ $(\widehat{K})^{2}$).

To illustrate the last statement we can consider an example from solid state
physics. Suppose that the system discussed in the beginning of this section
has locally a two dimensional character (i.e., locally $V_{phys}$ represents
a two-dimensional crystal). Then we can attempt to describe such a system in
terms of $R^{2}$. If we project $V_{phys}$ onto $R^{2}$, then we find that
in the case of an arbitrary topology of the two-dimensional crystal $%
V_{phys} $ the bias operator will have eigenvalues $N_{k}=0,1,2,...$. E.g.,
if $\widehat{K}$ is the diagonal in the position representation (i.e., $[%
\widehat{K}\widehat{X}]=0$), then $N_{k}$ is merely the number of different
points of the crystal (i.e., the number of two-dimensional sheets) which
correspond to the point $x_{k}\in R^{2}$. All such points however have
different positions in $R^{3}$, i.e., they differ in the extra coordinate $%
z_{k}^{a}$($a=1,2,...,N_{k}$) orthogonal to $R^{2}$. However, if the
Hamiltonian of the system does not include the extra coordinate $z_{k}$, it
is not measurable (without additional means) and states, which differ in the
extra coordinate only, become physically indistinguishable and
quasi-particles will obey a generalized statistics. In particular in the
given example $N_{k}$ gives the maximal number of electrons which can occupy
the same position $x_{k}\in R^{2}$. For more details see also Refs. \cite%
{K03,K04}.

In this manner we see that a non-trivial topological structure of the
physical space (as compared to the coordinate space) can indeed produce a
specific bias of all observables. We note that in this case the field theory
does not change at all, i.e., the mathematical structure of equations of the
motion (e.g., the Einstein equations) remains the same. What is actually
modified here is spatial properties of physical fields\footnote{%
In general physical fields should be understood as generalized fields Refs. 
\cite{K04}.} which are expressed by expansions of the type (\ref{bmod}). In
particular, every discrete point source (e.g., a galaxy or a star) is not
the Dirac delta function any more but acquires a specific distribution in
space (e.g., see (\ref{ds})) which reflects the topological structure of the
physical space (the density of degrees of freedom $N_{k}$).

\section{The bias function $b\left( r\right) $}

In what follows we, for the sake of simplicity, restrict our consideration
to the Newtonian limit (for the range of applicability of this limit see,
e.g., Ref. \cite{Peeb}). In a homogeneous and isotropic Universe the set of
solution (\ref{ss}) can be taken in the form $f_{k}=\left( 2\pi \right)
^{-3/2}g_{k}\left( t\right) e^{ikr}$ (i.e., states of particles can be
classified by wave numbers $k$), while the density of states $N_{k}$ is an
arbitrary function of $|k|$. If we assume that topology transformations have
stopped after the quantum period in the evolution of the Universe, then the
function $N_{k}$ will depend on time via only the cosmological shift of
scales, i.e., $k\left( t\right) \sim 1/a\left( t\right) $ (where $a\left(
t\right) $ is the scale factor). Thus, any point source undergoes the bias 
\begin{equation}
\delta \left( \vec{r}\right) \rightarrow \Delta \left( \vec{r},t\right) =%
\frac{1}{2\pi ^{2}}\int\limits_{0}^{\infty }\left( N_{k}k^{3}\right) \frac{%
\sin \left( kr\right) }{kr}\frac{dk}{k}.  \label{delta}
\end{equation}

The case of a simple topology corresponds to $N_{k}=1$, while in a
non-trivial case ($N_{k}-1\neq 0$) every point mass $M_{0}$ is surrounded
with an additional spherical "dark" halo 
\begin{equation}
\rho _{DM}\left( r,t\right) =M_{0}b\left( r,t\right) =\frac{M_{0}}{2\pi ^{2}}%
\int\limits_{0}^{\infty }\left( N_{k}\left( t\right) -1\right) k^{3}\frac{%
\sin \left( kr\right) }{kr}\frac{dk}{k}  \label{DM}
\end{equation}%
and the Newton's potential modifies as 
\begin{equation}
\phi =-\frac{GM_{0}}{r}\left( 1+f\left( r,t\right) \right) ,  \label{Np}
\end{equation}%
where the correction $f\left( r,t\right) $ relates to the bias function $%
b\left( r,t\right) $ according to ($f\left( r,t\right) ^{\prime }=\partial
f/\partial r$)%
\begin{equation}
b\left( r\right) =-\frac{f\left( r,t\right) ^{\prime \prime }}{4\pi r}.
\end{equation}%
Thus, the relation between visible matter $\rho _{vis}$ and DM is indeed
given by (\ref{br}) which in the Newtonian limit for the homogeneous and
isotropic Universe reduces to 
\begin{equation}
\rho _{DM}\left( \vec{r},t\right) =\widehat{B}\rho _{vis}=\int b\left(
\left\vert \vec{r}-\vec{r}^{\prime }\right\vert ,t\right) \rho _{vis}\left( 
\vec{r}^{\prime },t\right) dV^{\prime }.  \label{ddm}
\end{equation}

The explicit specification of the bias function $b\left( r,t\right) $ is, in
the first place, the problem of observational cosmology. Indeed, for Fourier
transforms there is a linear relation between DM and visible sources 
\begin{equation}
\rho _{DM}\left( \vec{k},t\right) =b\left( \vec{k},t\right) \ \rho
_{vis}\left( \vec{k},t\right)   \label{b}
\end{equation}%
which allows to find empirically the bias operator $\widehat{B}_{emp}$ (we
recall that the total source $\rho _{tot}=$ $\rho _{DM}+\rho _{vis}$ can be
restored from the measured spectrum of $\Delta T/T$ in CMB \cite{WMAP} and
the observed peculiar velocity field). It is quite obvious that such an
empirical bias operator $\widehat{B}_{emp}$ (in virtue merely of its
definition) describes perfectly DM effects at very large scales (where
inhomogeneities have the linear character). The nontrivial moment here is
that all theories which predict the same bias $b\left( r,t\right) $ for the
modern Universe are observationally indistinguishable (at least it requires
involving more subtle effects). We also note that in the more general case
the bias relations should be described by two functions $\rho _{DM}=b_{\rho
}\rho _{vis}$ and $p_{DM}=b_{p}p_{vis}$ (where $p$ is the pressure) which
for a homogeneous distribution reduce merely to functions of time $b_{\rho
,p}^{\prime }\left( t\right) $. Thus, the bias relation give the possibility
to account for Dark Energy as well (i.e., the observed\footnote{%
We point out however that the accelerated expansion can not be considered as
an established fact yet, for the presence of considerable uncertainties of a
theoretical character. } accelerated expansion of the Universe \cite{Ac}).

A specific feature of CDM models is that the relation between the two
sources appears as a result of the dynamics and, therefore, the effective
bias function $b\left( r,t\right) $ carries in general a nonlinear
character. The "great" success of CDM models in reproducing the large-scale
structure (LSS) of the Universe is somewhat exaggerated, for at very large
scales density perturbations are still at the linear stage of the
development and, therefore, the bias $b_{emp}\left( \vec{k},t\right) $
straightforwardly defines the set of appropriate initial conditions $%
b_{emp}\left( \vec{k},t\right) $ $=$ $D\left( t\right) b_{0}\left( \vec{k}%
\right) $ (where $b_{0}\left( k\right) =\rho _{DM}^{0}\left( \vec{k}\right)
/\rho _{vis}^{0}\left( \vec{k}\right) $ and $D\left( t\right) $ accounts for
the evolution of perturbations) depending on the exact behavior of the scale
factor $a\left( t\right) $. In this sense LSS alone in principle cannot
distinguish a model. On the contrary, at smaller scales (e.g., in galaxies
and clusters) perturbations are in a strongly nonlinear regime, the bias
operator $\widehat{B}$ acquires a nonlinear dependence on the distribution
of matter and CDM models fail \cite{Core,CE}.

Leaving the problem of the empirical determining of $\widehat{B}$ aside, in
what follows we consider a model expression for the bias $b\left( r\right) $

\begin{equation}
b\left( r\right) =\frac{1}{4\pi r_{0}r^{2}}\theta \left( r-r_{\max }\right)
,\   \label{K}
\end{equation}%
where $\theta \left( x\right) $ is the step function. $b\left( r\right) $
produces the correction to the Newton's potential (\ref{Np}) of the form%
\begin{equation}
f\left( r\right) =\left\{ 
\begin{array}{c}
r/r_{0}\ln \left( r_{\max }e/r\right) \,,~\ as\ r\leq r_{\max }, \\ 
r_{\max }/r_{0},\ as\ r>r_{\max }\ .%
\end{array}%
\right.  \label{f}
\end{equation}%
Such a bias was derived in Refs. \cite{KT02,K03} for the case of a
homogeneous and isotropic Universe under the assumption that the topological
structure (i.e., the number density of degrees of freedom $N_{k}$) of the
early Universe is described merely by the thermal equilibrium state\footnote{%
We note that the actual bias depends on the specific picture of topology
transformations in the early Universe and may differ from (\ref{K}).}.
Presumably, topology changes have occurred during the quantum stage of the
evolution of the Universe and at present are strongly suppressed. This means
that after the quantum period the topological structure remains constant.
Therefore, the isotropic cosmological expansion is accompanied only with the
cosmological shift of the parameters $r_{0}$ and $r_{\max }$ (i.e., $%
r_{0,\max }\left( t\right) =a\left( t\right) \widetilde{r}_{0,\max }$)
without any change in the form of the bias function (\ref{K}).

After the radiation dominated stage, however, the small initial adiabatic
perturbations (which are directly measured in CMB e.g., by WMAP \cite{WMAP})
start to grow and considerably shrink the Universe from galactic to
supercluster scales. The latter results in the further transformation of the
bias function $b\left( \left\vert x-x^{\prime }\right\vert \right)
\rightarrow b\left( \left\vert x-x^{\prime }\right\vert ,x^{\prime
},t\right) $. To derive rigorously the bias in a general inhomogeneous case
we have to construct a set of exact solutions to the wave equation (\ref{w})
which in turn depend on the distribution of matter and, therefore, on the
bias. In the simplest case however the inhomogeneity of the Universe can be
accounted for by an additional dependence of the parameters of the bias
function (\ref{K}) on the position in space. Indeed, the adiabatic growth of
density perturbations can be viewed as if the rate of the expansion were
different in different parts of the Universe $a\left( t\right) \rightarrow
a\left( t,x\right) $ which produces the respective shifts $r_{0,\max }\left(
t,x\right) \sim a\left( t,x\right) \widetilde{r}_{0,\max }$. Such an
additional shift is considerable indeed, e.g., the mean density of our
Galaxy has the order $\rho _{g}\sim 10^{6}\rho _{cr}$ (while the density
behaves as $\rho \sim 1/a^{3}$) and therefore for our Galaxy $r_{g0}$ should
be less in $10^{2}$ times, than the respective mean parameter $r_{0}$ for
the homogeneous Universe.

\section{The bias function and Dark Matter halos}

It is rather surprising that already the simplest function (\ref{K}) shows a
rather good qualitative agreement with the observed picture of the present
Universe. First of all it is consistent with the observed cored distribution
of DM in galaxies \cite{Core,CE}. Indeed, if $\rho _{vis}\left( r\right) $
is a rather smooth monotonously decreasing function of $r$, then from (\ref%
{K}) and (\ref{ddm}) we find that DM density reaches the maximal value in
the central region of a galaxy (i.e., as $r\ll R_{D}$, where $R_{D}$ has the
order of the stellar exponential scale length) 
\begin{equation}
\rho _{DM}\left( r\right) \simeq \rho _{DM}\left( 0\right) =\int \frac{1}{%
4\pi r_{0}r^{\prime 2}}\rho _{vis}\left( \vec{r}^{\prime },t\right)
dV^{\prime },  \label{max}
\end{equation}%
while for $r\gg R_{D}$ we find $\rho _{DM}\left( r\right) \simeq
M_{vis}/\left( 4\pi r_{0}r^{2}\right) $ (where $M_{vis}=\int \rho _{vis}dV$)
which can be combined by the interpolation formula%
\begin{equation}
\rho _{DM}\left( r\right) =\rho _{DM}\left( 0\right) \frac{R_{C}^{2}}{%
R_{C}^{2}+r^{2}},  \label{corem}
\end{equation}%
where $R_{C}^{2}=M_{vis}/\left( 4\pi r_{0}\rho _{DM}\left( 0\right) \right)
\simeq \alpha ^{2}R_{D}^{2}$ ) which explains the observed strong
correlation between $R_{C}$ and $R_{D}$ \cite{S04}. We note that the actual
value of the ratio $R_{C}$/$R_{D}=\alpha $ depends on the distribution of
the visible matter in a galaxy $\rho _{vis}\left( \vec{r},t\right) $ and the
definition of $R_{D}$ (e.g., if we assume in (\ref{max}) that $\rho _{vis}=%
\overline{\rho }$ within the ball $r<R_{D}$, then $\alpha ^{2}=1/3$).

The bias (\ref{K}) shows also that in the interval of scales $r<r_{\max }$
the dynamical mass of a point source increases with the radius as $%
M_{dyn}=M_{0}\left( 1+r/r_{0}\right) $, while for $r>r_{\max }$ it acquires
a new constant value $M_{\max }\sim M_{0}\left( 1+r_{\max }/r_{0}\right) $
and the ratio $r_{\max }/r_{0}$ defines the fraction of DM in the total
(baryons plus dark matter) density.

The minimal scale $r_{0}$ is different for different galaxies (i.e., $%
r_{0}=r_{0}\left( x\right) $ is a slow function of the position) and it has
the order $r_{0}\sim 1-5$ $kpc$ (it is the scale at which DM starts to show
up), while the value of $r_{\max }$ is not so well fixed by observations.
The analysis of the mass-to-light ratio $M/L$ shows that it increases with
scales for galaxies and groups but flattens eventually and remains
approximately constant for clusters (e.g., see Ref. \cite{B94}). This gives
an estimate $r_{\max }\gtrsim 1-5Mpc$ or $r_{\max }/r_{0}\gtrsim 10^{3}$.
Such a fraction of DM is indeed observed in LSB (Low Surface Brightness)
galaxies in which the ratio can reach $M/L\sim \left( 200-600\right)
M_{\odot }/L_{\odot }$. It however looks inconsistent with predictions of
CDM models and observed peculiar velocities in clusters which favor $\rho
_{DM}/\rho _{b}\sim 20$. The most drastic estimate comes from the observed
fractal distribution of galaxies which suggests $r_{\max }\gtrsim 200Mpc$
and $r_{\max }/r_{0}\gtrsim 10^{5}$ \cite{k02}. We however note that the
absolute boundary for $r_{\max }$ is given by the Hubble radius $r_{\max
}\leq R_{H}$ which gives $r_{\max }/r_{0}\leq R_{H}/r_{0}\sim 10^{6}-10^{7}$%
, while all values $r_{\max }\geq R_{H}$ are indistinguishable from
observations.

It turns out however that all those estimates are consistent with each other
and give only the lowest boundary for the DM fraction, for in any system
some essential portion of DM forms an inner core (i.e. the central constant
density region) and does not contribute to the local dynamics. Indeed, DM
consists of spherical halos (\ref{DM}) around every point source and,
therefore, the relationship between the baryon density and DM has a
non-local nature with the characteristic scale $r_{\max }$. The density of
DM in a point of space (and respectively the local dynamics) is formed by
all sources within the sphere of the radius $r_{\max }$ and it depends
essentially on the distribution of the sources. E.g., if we take $\rho
_{vis}\left( x,t\right) =\sum_{a}M_{a}\delta \left( R_{a}\right) $, then
from (\ref{ddm}) and (\ref{K}) we get for DM density 
\begin{equation}
\rho _{DM}\left( x,t\right) =\sum_{R_{a}\leq r_{\max }}\frac{M_{a}}{4\pi
r_{0}R_{a}^{2}}\geq \frac{r_{\max }}{r_{0}}\frac{\left\langle \rho
_{vis}\right\rangle }{3},  \label{DMD}
\end{equation}%
where $R_{a}=\left\vert x-x_{a}\left( t\right) \right\vert $ and $%
\left\langle \rho _{vis}\right\rangle =\sum_{R_{a}<r_{\max }}M_{a}/\left(
4/3\pi r_{\max }^{3}\right) $ is the mean density of the visible matter
within the sphere of the radius $r_{\max }$. For a uniform distribution of
matter this reads $\left\langle \rho _{DM}\right\rangle =\left( r_{\max
}/r_{0}\right) \left\langle \rho _{vis}\right\rangle $. From (\ref{DMD}) we
see that the DM density reaches the minimal possible value $1/3\left\langle
\rho _{DM}\right\rangle $ in the case when all sources are at the distance $%
R_{a}=r_{\max }$ (e.g., in the center point of a void), while according to (%
\ref{corem}) at a source $M_{a}$ it has a local maximum $\rho _{DM}\simeq
\left( \ell _{a}/r_{0}\right) \left\langle \rho _{a}\right\rangle /3$ (where 
$\ell _{a}$ is a characteristic dimension of the source and $\left\langle
\rho _{a}\right\rangle =3M_{a}/4\pi \ell _{a}^{3}$ ).

(\ref{DMD}) shows that DM halos smooth the observed strong inhomogeneity in
the distribution of baryons which considerably reduces the inhomogeneity in
the total density. By other words a considerable portion of DM acquires the
cored (\ref{corem}) (i.e., the quasi-homogeneous) character and switches off
from the local dynamics. This, in turn, leads to a renormalization of the
maximal scale $r_{\max }\rightarrow R_{\ast }$ in (\ref{K}) and, therefore,
changes the fraction of DM observed in a system $\rho _{DM}^{\prime }/\rho
_{b}\sim R_{\ast }/r_{0}$. In such a picture the scale $R_{\ast }$ is a
specific parameter of a system and this explains the small value for the
ratio $R_{\ast }/r_{0}$ observed in clusters.

Indeed, consider a group of galaxies of the characteristic dimension $L$.
Such a group can be characterized by the mean density $\left\langle \rho
_{DM}\right\rangle _{L}=1/L^{3}\int \rho _{DM}\left( x,t\right) d^{3}x$ and
perturbations $\delta _{DM}\left( x,t\right) =\rho _{DM}\left( x,t\right)
/\left\langle \rho _{DM}\right\rangle _{L}-1$. Near a particular galaxy in
the group ($r_{g}\left( t\right) =0$ and $M_{g}\ll \sum M_{a}$) we find from
(\ref{DMD}) 
\begin{equation}
\delta _{DM}\left( r\right) \simeq \frac{R_{\ast }^{2}}{r^{2}}-1,
\end{equation}%
where $R_{\ast }$ is the effective size $R_{\ast }$ of the DM halo 
\begin{equation}
\frac{R_{\ast }^{2}}{r_{0}^{2}}=\frac{M_{g}}{4\pi r_{0}^{3}\left\langle \rho
_{DM}\right\rangle _{L}}.  \label{R}
\end{equation}%
For $r>R_{\ast }$ we see that $\delta _{DM}<0$ and in the interval $%
L>r>R_{\ast }$ this function oscillates around the zero point (the exact
behavior depends on the distribution of galaxies in the group and is not
important).

The homogeneous background contributes only to the local Hubble flow which
can be accounted for by the expanding reference frame $x=a\left( t\right) r$
(e.g., see Ref. \cite{Peeb}). Thus, the actual Newton's potential of the
galaxy takes the form 
\begin{equation}
\delta \phi \left( r,t\right) =-Ga^{2}\left( \frac{M_{g}}{r}+\delta
F_{DM}\left( r,t\right) \right) ,\ 
\end{equation}%
with 
\begin{equation}
\delta F_{DM}\left( r,t\right) =\sum_{i}M_{i}\frac{f\left( \left\vert
r-r_{i}\left( t\right) \right\vert \right) }{\left\vert r-r_{i}\left(
t\right) \right\vert }+2/3\pi \left\langle \rho _{DM}\right\rangle
_{L}r^{2}=M_{g}\frac{f\left( r,R_{\ast }\right) }{r}+\mu \left( r,t\right)
\label{FD}
\end{equation}%
where we subtracted the homogeneous component $\delta \phi =\phi
-\left\langle \phi \right\rangle _{L}$ (with $\left\langle \phi
\right\rangle _{L}=2/3\pi Ga^{2}\left\langle \rho \right\rangle _{L}r^{2}$) 
\cite{Peeb}, $\mu \left( r,t\right) $ accounts for variation of $\delta
_{DM} $ for $r>R_{\ast }$, and $f\left( r,R_{\ast }\right) $ is given by (%
\ref{f}) with the replacement $r_{\max }\rightarrow R_{\ast }$. The function 
$\delta F_{DM}$ defines the contribution of the DM halo and we recall that
the use \textit{of the empirical bias function} $b_{emp}\left( r,r^{\prime
},t\right) $ (or equivalently $f\left( r.t\right) $) \textit{automatically
reproduces all actual DM halos in astrophysical systems. }

Thus, we see that near a source the function $\delta F_{DM}$ has the
logarithmic behavior\footnote{%
We note that in the presence of a continuous medium (e.g., of gas) the
behavior may essentially change.}. At the distance $R_{\ast }$ the logarithm
switches off and the ratio $R_{\ast }/r_{0}$ defines the maximal value for
the DM mass in a galaxy or a cluster which can be observed from the local
dynamics. We recall that the value $r_{0}$ is different for different
galaxies. In addition to this fact the expression (\ref{R}) shows the
general tendency that the ratio $R_{\ast }/r_{0}$ (and therefore the maximal
discrepancy between the dynamical mass and luminous matter) is smaller in
high density regions of space and larger in low density regions. This
qualitative feature agrees with discrepancies observed in LSB and HSB
galaxies.

\section{The background distribution of baryons and $r_{\max }$}

Consider now properties of the homogeneous and isotropic background. In the
standard models there exist the only case which corresponds to the
homogeneous distribution of baryons. If we accept the bias of wave equations
(\ref{M}), there appears a new possibility. Indeed, the homogeneity of the
Universe (or the cosmological principle) requires the total distribution of
matter (baryons plus dark halos) to be homogeneous, while properties of the
baryon distribution are not fixed well. The latter may have a quite
irregular character. Exactly such a situation takes place in the case of a
fractal distribution of baryons. Consider a sphere of a radius $r$. Then the
total mass within the radius $r$ is given by 
\begin{equation}
M_{tot}\left( r\right) \simeq m_{b}\left( 1+r/r_{0}\right) N_{b}\left(
r\right) +\delta M\left( r\right) ,  \label{ch}
\end{equation}%
where $N_{b}\left( r\right) $ is the actual number of baryons, $m_{b}$ is
the baryon mass, and $\delta M\left( r\right) $ accounts for corrections
and, in particular, for the contribution of dark halos of baryons\ from the
outer region. The homogeneous distribution means that the total mass behaves
as $M_{tot}\left( r\right) =\left\langle \rho \right\rangle V\left( r\right)
\sim r^{3}$. And for $r\gg r_{0}$ this can be reached by the fractal law $%
N\left( r\right) \sim r^{D}$ with $D\approx 2$ (the exact equality cannot be
reached, for the presence of the additional term $\delta M\left( r\right) $%
). Such a law works up to the scale $r_{\max }$ upon which the distribution
of baryons crosses over to homogeneity\footnote{%
The distribution of stars in galaxies shows also a fractal behavior. In this
sense we can say that the fractal law forms the cored distribution (\ref%
{corem}) with $R_{C}\sim R_{H}$.}.

There exists at least two strong arguments in favor of the fractal
distribution of baryons. The first argument is that the fractal distribution
is more stable gravitationally. Indeed, let us fix the total density $\Omega
_{tot}=\rho _{tot}/\rho _{cr}\sim 1$ (where $\rho _{cr}$ is the critical
density) and the baryon fraction $\rho _{b}/\rho _{tot}\sim r_{0}/r_{\max }$%
. In the case of the fractal distribution this fraction reaches only at
scales $r\geq r_{\max }$, while at smaller scales baryons are distributed
rather irregularly.

Consider first a homogeneous distribution of baryons. Now if we consider a
small displacement of a particular baryon (or of a homogeneous group of
baryons), then such a displacement will produce the same displacement of the
dark halo (attached to the baryon). So the resulting perturbation increases
in $r_{\max }/r_{0}$ times. The maximal scale $r_{\max }$ should be larger
than $100-200Mpc$, and therefore the increase should be more than $%
10^{5}-10^{6}$. In the primordial plasma the domination of radiation prevent
the growth of perturbations in the gravitational potential and, therefore,
such fluctuations are strongly suppressed. However, there also do exist
collective fluctuations in the density of baryons which do not affect the
metric perturbations\footnote{%
The presence of metric perturbations at some level $\Delta \rho _{tot}/\rho
_{tot}$ $\sim $ $10^{-5}$ is essential however, otherwise the fractal
structure in baryonic matter will not form. For fluctuations the bias
relation reads $\Delta \rho _{DM}=B\Delta \rho _{vis}-\left\langle B\Delta
\rho _{vis}\right\rangle =F\Delta \rho _{vis}$, which defines a new operator 
$F$. Thus, fluctuations which obey $\left( F-1\right) \Delta \rho
_{vis}\approx 0$ do not affect the metric and $\Delta \rho _{tot}\approx
const$.} and the total density of matter. According to (\ref{M}) such
fluctuations do not affect the total (effective) charge density and,
therefore, the radiation dominated stage cannot prevent a specific
redistribution of baryons. By other words perturbations of such a type could
increase long before the recombination. They do not change the total density 
$\delta \rho _{tot}=\delta \rho _{b}+\delta \rho _{DM}=const$ and can be
called compensational sound waves. In the very early Universe high
temperatures transform baryons from the more constrained state which
corresponds to a homogeneous distribution of baryons to the less constrained
and more stable state which corresponds to the fractal distribution. We
note, however, that during the radiation dominated stage when $\delta \rho
_{b}+\delta \rho _{DM}\approx 0$ and $\Omega _{b}+\Omega _{DM}\sim 1$
perturbations in the baryon number density cannot grow to an arbitrary large
value, but are restricted by $\Omega _{b}\leq 1$, (i.e., $\delta \rho
_{b}/\rho _{b}\leq \rho _{tot}/\rho _{b}\sim r_{\max }/r_{0}$).

Consider now the case of the fractal distribution. According to (\ref{ch})
the fractal distribution of dark halos forms the homogeneous background of
the total density. Now any small displacement of a baryon does not change
the character of the dark halos distribution and, therefore, the increase is
essentially suppressed ($r_{\max }/r_{0}\rightarrow R_{\ast }/r_{0}$). By
other words \textit{the stable equilibrium distribution can be defined as
such a distribution of baryons for which perturbations in the baryon density
produce the minimal response in the total density}. The bias of the
electromagnetic field (\ref{M}) insures the absence of strong fluctuations
in the CMB temperature caused by the fractal distribution of baryons. This
may be used to estimate the value of the fraction $r_{\max }/r_{0}$.

Indeed, the first estimate comes from the upper boundary for the scale of
the cross-over to the homogeneity in the observed galaxy distribution $%
r_{\max }\geq 100-200Mpc$ which gives $r_{\max }/r_{0}\geq 10^{5}$. From the
other side, the observed CMB gives $\Delta T/T$ $=$ $\frac{1}{3}\Delta \rho
_{tot}/\rho _{tot}$ $\sim $ $10^{-5}$ at the moment of recombination, and
the fractal distribution causes perturbations in the total density $\Delta
\rho _{tot}\sim \left( R_{\ast }/r_{0}\right) \rho _{b}$ (where the factor $%
R_{\ast }/r_{0}$ appears as the contribution from dark halos) and therefore $%
\Delta \rho _{tot}/\rho _{tot}\geq \left( R_{\ast }/r_{0}\right) \rho
_{b}/\rho _{tot}$ $\sim R_{\ast }/r_{\max }\leq 10^{-5}$. We see that both
estimates agree and give $r_{\max }/r_{0}\geq \left( R_{\ast }/r_{0}\right)
10^{5}$. As it was shown above the ratio $R_{\ast }/r_{0}$ takes the minimal
value for the equilibrium fractal distribution. So that the value $r_{\max }$
(which is the scale of the cross-over to the homogeneity in the visible
matter) will increase, if at the moment of the recombination the ideal
fractal distribution had not been achieved yet.

The second argument is based on a more correct interpretation of the dark
matter effects. Indeed, the bias of the wave equation (\ref{M}) should be
understood as the fact that at large scales our Universe possesses a rather
unusual geometric (or topological) properties. These geometric properties
are reflected in the behavior of the Green function (\ref{M}) which for $%
r>r_{0}$ acquires effectively the two-dimensional character (e.g., for $%
N_{k}\sim 1/\left( kr_{0}\right) $ we get $G\left( r,\tau \right) \sim \frac{%
1}{rr_{0}}\ln \left( \tau -r\right) /\left( \tau +r\right) $) and,
therefore, such a geometry should be reflected in the distribution of matter
(sources). By other words at scales $r>r_{0}$ our Universe acquires an
effective dimension $D\approx 2$ (e.g., see Ref. \cite{K03}) which explains
the two-dimensional character of the spatial distribution of baryons. By
other words we may imagine that our Universe represents a fractal (the space
is \textquotedblright more dense\textquotedblright\ on a fractal set than
outside (e.g., see Ref. \cite{K03})) and within such a fractal the matter
has a homogeneous distribution. In such a picture the fractal distribution
is the only thermal equilibrium state. We note that in the case $r_{\max
}<\infty $ such a state can never be utterly homogeneous but always includes
equilibrium fluctuations of the order $\Delta \rho _{tot}/\rho _{tot}\sim
r_{0}/r_{\max }$.

\section{Variation of interaction constants}

In the present section we show that the structure formation in the present
Universe leads to a specific variation with time of all interaction
constants. As an example we consider the variation of the gravitational
constant. Indeed, the cosmological evolution is described by the scale
factor $a\left( t\right) $ which obey the equation \cite{Peeb} (we consider
the case $p=0$) 
\begin{equation}
\frac{d^{2}a}{dt^{2}}=-\frac{4\pi G}{3}\left\langle \rho _{tot}\right\rangle
a=-\frac{4\pi G}{3}\left( 1+\frac{r_{\max }}{r_{0}}\right) \left\langle \rho
_{vis}\right\rangle a.  \label{FE}
\end{equation}%
This equation can be interpreted as if the gravitational constant
renormalizes as $G\rightarrow \widetilde{G}=G\left( 1+r_{\max }/r_{0}\right) 
$ (we recall that in the inhomogeneous case it depends on scales as well).

The mean density of the visible matter behaves as $\left\langle \rho
_{vis}\right\rangle $ $\sim 1/a^{3}$. Thus, the evolution of the scale
factor $a\left( t\right) $ depends essentially on the behavior of the ratio $%
r_{\max }/r_{0}$. During the radiation dominated stage $\left\langle \rho
_{tot}\right\rangle =\left\langle \rho _{\gamma }\right\rangle $, the growth
of density perturbations is suppressed and therefore the bias function (\ref%
{K}) in the commoving frame (i.e., in the expanding reference system $x=ar$)
does not change. Thus, during the radiation dominated stage the ratio $%
r_{\max }/r_{0}=const$. This remains true and on the subsequent stage, while
inhomogeneities in the total density remain small $\delta =\Delta \rho
_{tot}/\rho _{tot}\ll 1$. The situation changes drastically when the
inhomogeneities reach the value $\delta \sim 1$. Upon this moment the time
shifts of the two scales $r_{0}$ and $r_{\max }$ disagree. Small scale
inhomogeneities develop first and switch off from the Hubble expansion. This
leads to the monotonic increase of the effective gravitational constant $%
\widetilde{G}$, i.e., of the ratio $r_{\max }/r_{0}\sim a^{\beta }$, which
gives for the matter density $\left\langle \rho _{tot}\right\rangle \sim
a^{\beta -3}$. While inhomogeneities remain small $\delta \leq 1$, both
scales increase with time as $r_{0}$, $r_{\max }\sim a$, and the exponent $%
\beta \sim 0$. When $\delta $ reaches the value $\delta \gtrsim 1$ the scale 
$r_{0}$ starts to collapse (galaxies start to form), while $r_{\max }$ is
still increasing $r_{\max }\sim a$. This leads to the fact that the exponent
becomes $\beta >1$ and DM behaves as "Dark Energy", e.g., in the case $\beta
=3$ DM behaves as the negative Lambda term $\Lambda =-4\pi G\left\langle
\rho _{DM}\right\rangle =const$. This kind of regime ends either when the
collapse of the scale $r_{0}$ ends (galaxies have stabilized and $%
r_{0}\simeq const$ and $\widetilde{G}\sim a$), or when the maximal scale $%
r_{\max }$ sufficiently deviates from the Hubble law $r_{\max }\sim a$.

The behavior of the minimal scale $r_{0}$ follows the local dynamics and can
be estimated as $r_{0}\sim \delta _{0}^{-1/3}a\widetilde{r}_{0}$, where $%
\delta _{0}$ is the mean perturbation within the radius $r_{0}$ and the
parameter $\widetilde{r}_{0}=const$. Analogously, the maximal scale is given
by $r_{\max }\sim \delta _{\max }^{-1/3}a\widetilde{r}_{\max }$, which gives 
$r_{\max }/r_{0}\sim \left( \delta _{\max }\left( t\right) /\delta
_{0}\left( t\right) \right) ^{-1/3}$ and therefore the effective
gravitational constant depends on time as $\widetilde{G}\left( t\right)
\approx G\left( 1+\left( \delta _{\max }\left( t\right) /\delta _{0}\left(
t\right) \right) ^{-1/3}\right) $.

The fact that the bias operator reflects the topological structure of space
means that all interaction constants undergo an additional renormalization
(e.g., see Ref. \cite{K04}) and acquire the same dependence on time. E.g.,
the fine structure constant takes the form $\widetilde{\alpha }\left(
k,t\right) =b\left( k,t\right) \alpha $ which gives for homogeneous fields $%
\widetilde{\alpha }\left( t\right) \approx \alpha \left( 1+\left( \delta
_{\max }\left( t\right) /\delta _{0}\left( t\right) \right) ^{-1/3}\right) $%
. It is remarkable that a small variation of the fine structure constant
seems to be observed at high red shifts \cite{R11}.

In conclusion of this section we note that the decrease of the scale $%
r_{0}\left( t\right) $ during the structure formation can also be used to
explain the apparent acceleration of the Universe which seems to be required
by observations of the type Ia supernovae \cite{Ac}. Indeed, according to (%
\ref{M}), (\ref{delta}), and (\ref{K}) at large distances $r\gg r_{0}$ the
Green functions behave as $G\sim 1/r_{0}$ and therefore the apparent
luminosity will also behave as $L\sim L_{0}/r_{0}$. Thus, the decrease of
the scale $r_{0}$ will formally look as a very strong evolutionary effect $E=%
\dot{L}/L\sim -\dot{r}_{0}/r_{0}>0\,$, which produces correction $%
q\rightarrow $ $q^{eff}=q-E/H$, e.g., see Ref. \cite{W} (where $H=\dot{a}/a$
and $q=-d^{2}a/dt^{2}/\left( aH^{2}\right) $). Thus, the observed
acceleration $q<0$ may merely mean nothing but the strong evolutionary
effect caused by the variation of $r_{0}$.\bigskip

\section{Conclusion}

In conclusion we briefly repeat basic results. First of all from the
observed strong dark-to-luminous matter coupling \cite{S04,Core,CE} we
derive the existence of a bias relation $T_{\mu \nu }^{DM}=F_{\mu \nu
}\left( T_{\alpha \beta }^{vis}\right) $ which allows us to re-write the
Einstein equations in the equivalent biased form $R_{\mu \nu }-\frac{1}{2}%
g_{\mu \nu }R=8\pi G\left( T_{\mu \nu }+F_{\mu \nu }(T_{\alpha \beta
})\right) $. The biased Einstein equations straightforwardly predict the
presence of a specific correction to the Newton's potential for a point
source $\phi =-GM\left( 1/r+F(r,t)\right) $.

The bias may have an arbitrary nature, CDM, MOND, any modification of
gravity, etc., which does not change the phenomenological results of this
paper. We however have suggested the bias which naturally appears in the
case when the topological structure of the actual Universe at very large
distances does not match properly that of the Friedman space (the open,
flat, or closed model). In that case not only the gravitational potential
but also all other physical fields undergo the bias and display some
discrepancy (i.e., the presence of DM halos around every point source $%
\delta \left( x-x^{\prime }\right) \rightarrow \Delta (x-x^{\prime })$).

In the linear approximation the bias relation $\rho _{DM}$ $=\widehat{B}\rho
_{vis}$ is described by the function $b\left( r,r^{\prime },t\right) $ (the
kernel of the bias operator) which admits the empirical definition. Then $%
b_{emp}\left( r,r^{\prime },t\right) $ (or equivalently its spectral
components $b\left( \vec{k},t\right) $) gives a rather simple tool for
confronting a theory of the structure formation with observations. Any
acceptable theory has to reproduce in details the specific form of the bias
function $b_{emp}$.

We have demonstrated that a specific choice of the bias (\ref{K}) $%
b=1/\left( 4\pi r_{0}r^{2}\right) \theta \left( r-r_{\max }\right) $ (which
is predicted by topology changes in the early Universe \cite{KT02,K03})
shows quite a good agreement with the observed picture of the modern
Universe (e.g., the fractal distribution of galaxies, cored DM distribution
in galaxies and rich clusters, variety of DM halos, etc.). It however
considerably changes the estimate for the mean density of baryons $%
\left\langle \rho _{DM}\right\rangle /\ \left\langle \rho
_{vis}\right\rangle \sim r_{\max }/r_{0}$ (this in turn is not in a conflict
with observations, for in the standard models the estimate $\Omega _{b}\sim
0.05$ is model dependent and uses essentially the idea of the homogeneous
distribution of baryons).

Finally, we have shown that the galaxy formation process causes a decrease
of the minimal scale $r_{0}\left( t\right) $ (and the increase of the ratio $%
r_{\max }/r_{0}$) and this gives rise to a specific dependence on time for
all interaction constants. In particular, this may give an explanation to
the observed variation (a small increase) in the fine structure constant 
\cite{R11}.

\bigskip

In conclusion I would like to acknowledge discussions with D. Turaev and M.
Milgrom on problems of modified gravity which inspired me for this research.
I indebted to P. Salucci for attracting my attention to the recent
observational results in DM physics \cite{S04,Core,CE}. I would also like to
acknowledge the hospitality of D. Turaev and the Center for Advanced Studies
in the Ben Gurion University of the Negev where this research has been
started.

\end{document}